\documentstyle[epsfig]{mn}

\newcommand{\gtrsim}{\mathrel{\hbox{\rlap{\lower.55ex \hbox {$\sim$}}
                   \kern-.3em \raise.4ex \hbox{$>$}}}}
\newcommand{\lesssim}{\mathrel{\hbox{\rlap{\lower.55ex \hbox {$\sim$}}
                   \kern-.3em \raise.4ex \hbox{$<$}}}}

\newif\ifAMStwofonts

\ifoldfss
  \ifCUPmtlplainloaded \else
    \NewTextAlphabet{textbfit} {cmbxti10} {}
    \NewTextAlphabet{textbfss} {cmssbx10} {}
    \NewMathAlphabet{mathbfit} {cmbxti10} {} % for math mode
    \NewMathAlphabet{mathbfss} {cmssbx10} {} %  "   "    "
  \fi
  \ifAMStwofonts
    \ifCUPmtlplainloaded \else
      \NewSymbolFont{upmath} {eurm10}
      \NewSymbolFont{AMSa} {msam10}
      \NewMathSymbol{\upi}     {0}{upmath}{19}
      \NewMathSymbol{\umu}     {0}{upmath}{16}
      \NewMathSymbol{\upartial}{0}{upmath}{40}
      \NewMathSymbol{\leqslant}{3}{AMSa}{36}
      \NewMathSymbol{\geqslant}{3}{AMSa}{3E}

    \fi
  \fi
\fi % End of OFSS

\ifnfssone
  \newmathalphabet{\mathit}
  \addtoversion{normal}{\mathit}{cmr}{m}{it}
  \addtoversion{bold}{\mathit}{cmr}{bx}{it}
  \newmathalphabet{\mathbfit} % math mode version of \textbfit{..}
  \addtoversion{normal}{\mathbfit}{cmr}{bx}{it}
  \addtoversion{bold}{\mathbfit}{cmr}{bx}{it}
  \newmathalphabet{\mathbfss} % math mode version of \textbfss{..}
  \addtoversion{normal}{\mathbfss}{cmss}{bx}{n}
  \addtoversion{bold}{\mathbfss}{cmss}{bx}{n}
  \ifAMStwofonts
    \ifCUPmtlplainloaded \else
      \UseAMStwoboldmath
      \makeatletter
      \new@mathgroup\upmath@group
      \define@mathgroup\mv@normal\upmath@group{eur}{m}{n}
      \define@mathgroup\mv@bold\upmath@group{eur}{b}{n}
      \edef\UPM{\hexnumber\upmath@group}
      \new@mathgroup\amsa@group
      \define@mathgroup\mv@normal\amsa@group{msa}{m}{n}
      \define@mathgroup\mv@bold\amsa@group{msa}{m}{n}
      \edef\AMSa{\hexnumber\amsa@group}
      \makeatother
      \mathchardef\upi="0\UPM19
      \mathchardef\umu="0\UPM16
      \mathchardef\upartial="0\UPM40
      \mathchardef\leqslant="3\AMSa36
      \mathchardef\geqslant="3\AMSa3E
    \fi
  \fi
\fi % End of NFSS release 1
\ifnfsstwo
  \DeclareMathAlphabet{\mathbfit}{OT1}{cmr}{bx}{it}
  \SetMathAlphabet\mathbfit{bold}{OT1}{cmr}{bx}{it}
  \DeclareMathAlphabet{\mathbfss}{OT1}{cmss}{bx}{n}
  \SetMathAlphabet\mathbfss{bold}{OT1}{cmss}{bx}{n}
  \ifAMStwofonts
    \ifCUPmtlplainloaded \else
      \DeclareSymbolFont{UPM}{U}{eur}{m}{n}
      \SetSymbolFont{UPM}{bold}{U}{eur}{b}{n}
      \DeclareSymbolFont{AMSa}{U}{msa}{m}{n}
      \DeclareMathSymbol{\upi}{0}{UPM}{"19}
      \DeclareMathSymbol{\umu}{0}{UPM}{"16}
      \DeclareMathSymbol{\upartial}{0}{UPM}{"40}
      \DeclareMathSymbol{\leqslant}{3}{AMSa}{"36}
      \DeclareMathSymbol{\geqslant}{3}{AMSa}{"3E}
    \fi
  \fi
\fi % End of NFSS release 2

\ifCUPmtlplainloaded \else
  \ifAMStwofonts \else % If no AMS fonts
    \def\upi{\pi}
    \def\umu{\mu}
    \def\upartial{\partial}
  \fi
\fi
\title[Radio observation of A0620$-$00]
  {Multiple ejections during the 1975 outburst of
A0620$-$00}

\author[Kuulkers et al.]
  {Erik Kuulkers$^1$
\thanks{E-mail: E.Kuulkers@sron.nl (EK),
  rpf@astro.uva.nl (RF), res@jb.man.ac.uk (RS), rjd@jb.man.ac.uk (RD),
im@jb.man.ac.uk (IM).},
  R.~P.~Fender$^{2\star}$, Ralph E.~Spencer$^{3\star}$, 
  Richard J.~Davis$^{3\star}$ and \cr Ian Morison$^{3\star}$\\
  $^1$Space Research Organization Netherlands, Sorbonnelaan 2, 3584 CA Utrecht,
  \&\ Astronomical Institute, Utrecht University, \\P.O.~Box 80000, 3507 TA
  Utrecht, The Netherlands\\
  $^2$Astronomical Institute ``Anton Pannekoek", University of Amsterdam and
  Center for High-Energy Astrophysics, Kruislaan 403, \\1098 SJ Amsterdam,
  The Netherlands\\
  $^3$University of Manchester, Nuffield Radio Astronomy Labs., Jodrell Bank,
  Macclesfield, Cheshire SK 11 9DL, United Kingdom}

\date{Accepted. Received.}
\pagerange{\pageref{firstpage}--\pageref{lastpage}}
\pubyear{1999}

\begin{document}

\label{firstpage}

\maketitle

\begin{abstract}

The well-known black-hole X-ray transient A0620$-$00 was a bright
radio source during the first part of its outburst in 1975. We have
revisited the available data and find for the first time evidence that
the source exhibited multiple jet ejections. Rapid radio spectral
changes indicate the addition of at least three new components which
are initially optically thick.  From single baseline interferometry
taken about three weeks after the start of the X-ray outburst we find
that the source is extended on arcsec scales and infer a relativistic
expansion velocity. Some of the other (soft) X-ray transients, such as
GS\,1124$-$68 and GS\,2000+25, show very similar X-ray outburst light curve
shapes to that of A0620$-$00, 
while their radio outburst light curve shapes are different. 
We suggest that this is due to the radio emission
being strongly beamed in outburst, whereas the X-ray emission remains 
isotropic. Since this effect is stronger at higher jet velocities,
this strengthens our conclusion that the jets in A0620$-$00 and other
soft X-ray transients move with relativistic speeds.

\end{abstract}

\begin{keywords}
accretion, accretion disks -- binaries: close -- stars: individual: A0620$-$00
-- X-rays: stars
\end{keywords}

\section{Introduction}

A0620$-$00, a low-mass X-ray binary black-hole transient, was
discovered in outburst almost 25 years ago.  It was detected by the
Sky Survey Experiment onboard Ariel V on August 3rd 1975 (Elvis et
al.\ 1975), and subsequently at various other wavelengths (see
Kuulkers 1998 for a recent review).  The radio source associated with
the (soft) X-ray transient was detected almost two weeks after the
start of the outburst and was visible for about a fortnight (e.g.\
Davis et al.\ 1975; Owen et al.\ 1976).  The initial model to explain
the radio outburst light curve was the synchrotron `bubble' model
(homogeneous adiabatically expanding sphere of relativistic electrons;
e.g.\ van der Laan 1966; Hjellming \&\ Han 1995).  However, this model
often does not describe those radio transient light curves which have
sufficient coverage (see e.g.\ Ball 1994), and the reality of
relativistic ejections from X-ray binaries is in all likelihood
considerably more complex.

Since we have now a bigger sample of (black-hole) radio transients, in
several of which the radio emission following X-ray outburst has been
clearly resolved into relativistic outflows (e.g.\ Mirabel \& Rodriguez
1994; Tingay et al.\ 1995; Hjellming \& Rupen 1995; Mioduszewski et
al.\ 1998), we decided to re-investigate the radio outburst of
A0620$-$00.  We report here on the outburst light curves, spectral
evolution, and a comparison with its X-ray outburst. We find
some evidence from single-baseline interferometry
for expansion of the source to an angular size of a several
arcsec. We also discuss the radio
observations in the framework of the radio transient sample for the
black-hole X-ray transients which have similar X-ray properties 
to A0620$-$00 and have been reasonably well covered in the radio, 
i.e.\ GS\,1124$-$68 and GS\,2000+25.

\begin{figure*}
\centerline{\hbox{
\psfig{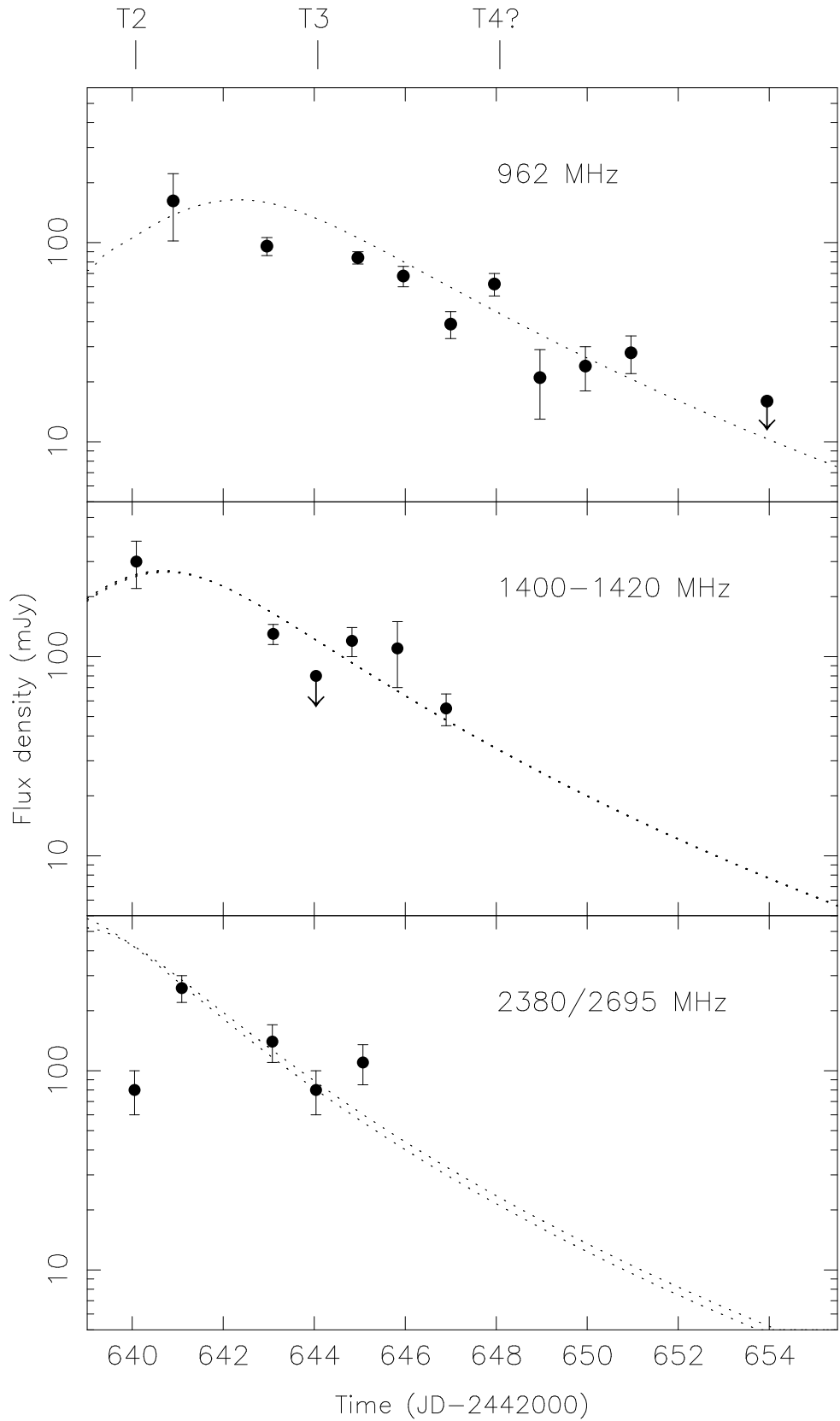}}
}
\caption{Radio light curves during the outburst of A0620$-$00 at three
frequencies, i.e.\ 962\,MHz (top), 1400--1420\,MHz (middle) and
2380/2695\,MHz (bottom). Upper limits are indicated by an arrow.
This figure is an update of the light curves
presented by Hjellming et al.\ (1988) and Hjellming \&\ Han (1995).
Also indicated are simplistic single synchrotron bubble fits which are
clearly inadequate representations of the data.
At the top we indicate the approximate times of the start of 
possible jet-ejections, see text.}
\end{figure*}

\newpage

\section{Radio emission from A0620$-$00}

\subsection{Observations}

\begin{table*}
\begin{center}
\begin{tabular}{lrrrl}
\multicolumn{5}{l}{{\bf Table 1}: Radio observations of A0620$-$00} \\
\hline
\multicolumn{1}{l}{Time} & \multicolumn{1}{c}{Frequency} &
\multicolumn{1}{c}{Flux density} & \multicolumn{1}{c}{Error} &
\multicolumn{1}{l}{Reference} \\
\multicolumn{1}{l}{(JD$-$2442000)} & \multicolumn{1}{c}{(MHz)} &
\multicolumn{1}{c}{(mJy)} & \multicolumn{1}{c}{(mJy)} &
\multicolumn{1}{l}{~} \\
\hline
636.5--639.5 &   962 & $<$5000 &    & Davis et al.\ 1975 \\
640.05       &  2380 &      80 & 20 & Craft 1975; Craft \&\ Davis 1998 \\
640.09       &  1400 &     300 & 80 & Owen et al.\ 1976 \\
640.1        & 10600 &  $<$250 &    & Feldman 1975 \\
640.90       &   962 &     162 & 60 & Davis et al.\ 1975; this paper \\
641.09       &  2695 &     260 & 40 & Owen et al.\ 1976 \\
642.052      &   151 &  $<$250 &    & Davis et al.\ 1975 \\
642.96       &   962 &      96 & 10 & Davis et al.\ 1975; this paper \\
643.08       &  2695 &     140 & 30 & Owen et al.\ 1976 \\
643.1        &  1418 &     130 & 15 & Owen et al.\ 1976 \\
644.04       &  2380 &      80 & 20 & Craft 1975; Craft \&\ Davis 1998 \\
644.04       &  1420 &   $<$80 &    & Craft 1975; Craft \&\ Davis 1998 \\
644.4        &   408 &     310 & 50 & Little et al.\ 1976 \\
644.83       &  1408 &     120 & 20 & Lequeux 1975 \\
644.96       &   962 &      84 &  6 & Davis et al.\ 1975; this paper \\
645.017      &   151 &  $<$250 &    & Davis et al.\ 1975 \\
645.07       &  2695 &     110 & 25 & Owen et al.\ 1976 \\
645.83       &  1408 &     110 & 40 & Lequeux 1975 \\
645.96       &   962 &      68 &  8 & Davis et al.\ 1975; this paper \\
646.017      &   151 &  $<$250 &    & Davis et al.\ 1975 \\
646.9        &  1420 &      55 & 10 & Owen et al.\ 1976 \\
647.00       &   962 &      39 &  6 & Davis et al.\ 1975; this paper \\
647.822      &  5000 &      20 &  7 & Scott 1975; Scott 1998 \\
647.96       &   962 &      62 &  8 & Davis et al.\ 1975; this paper \\
647.998      &   151 &  $<$250 &    & Davis et al.\ 1975 \\
648.822      &   151 &  $<$250 &    & Davis et al.\ 1975 \\
648.87       &  4600 &      28 &  6 & Bieging \&\ Downes 1975 \\
648.96       &   962 &      21 &  8 & Davis et al.\ 1975; this paper \\
649.96	     &   962 &      24 &  6 & Davis et al.\ 1975; this paper \\
650.96       &   962 &      28 &  6 & Davis et al.\ 1975; this paper \\
651.95       &   962 &   $<$16 &    & Davis et al.\ 1975; this paper \\
661.5--662.5 &   408 &   $<$10 &    & Little et al.\ 1976 \\
662.5        &  4600 &   $<$20 &    & Bieging \&\ Downes 1975 \\
663.5        &  4600 &   $<$20 &    & Bieging \&\ Downes 1975 \\
672.5        &   408 &   $<$10 &    & Little et al.\ 1976 \\
774.5        &   408 &   $<$10 &    & Little et al.\ 1976 \\
\hline
\end{tabular}
\end{center}
\end{table*}

We have collected all available radio observations of A0620-00 during
its 1975 outburst as reported in the literature.  For convenience we give these
observations in chronological order in Table 1. Note that for some of
the radio observations no exact times were available in the
literature; these have been updated by us from private communications.
The measurements by Davis et al.\ (1975) were used after a reassessment 
of errors.

In the text we will refer to time as given by JD$-$2442000.

We note that, apart from an account of the radio observations of A0620$-$00,
Lequeux (1975) mentions that `there is a 0.33\,Jy radio source
following by 27\,s'. At the declination of A0620-00 this implies an
angular separation of $\sim$6.8 arcmin; as such this second, bright
radio source is unlikely to be related to A0620-00.
Inspection of the NRAO VLA Sky Survey archive 
(NVSS; Condon et al.\ 1998) 
reveals several sources at about the correct location and flux densities; 
Lequeux's following source is in all likelihood then this group of relatively
unvarying field objects.

\subsection{Light curves and spectral evolution}

In Fig.~1 we present the radio outburst light curves of
A0620$-$00 at frequencies of 962\,MHz, 1400--1420\,MHz and at
2380/2695\,MHz. This is an update of figure~2 of Hjellming
et al.\ (1988; see also Hjellming \&\ Han 1995). 
Together with the data points we also show
synchrotron bubble model light curves as given by them at the various
frequencies\footnote{An error is present in the equation (2) of
Hjellming et al. (1988);
however, the model light curve plots are good. For the correct equation
we refer to e.g.\ Hjellming \&\ Johnston (1988) and Ball et al.\
(1995). Note also that the frequencies labeling the model light curves
in Hjellming et al. (1988) are wrong; they should read from top to bottom: 
2.7\,GHz, 1.4\,GHz and 0.96\,GHz.}.

\begin{figure}
\centerline{
\epsfig{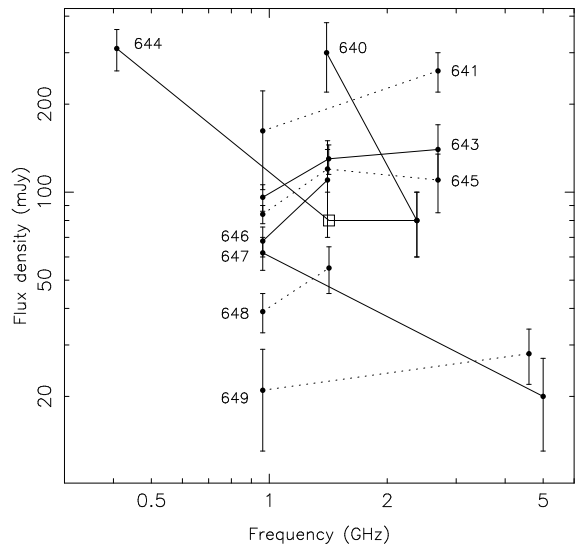}
%bbllx=21pt,bblly=65pt,bburx=567pt,bbury=707pt,width=8.5cm}
}

\caption{Radio spectral evolution of A0620$-$00 as a function of time.
The spectra are constructed from data points which were obtained within
a maximum time span of 0.36~days. They have been connected to guide
the eye. For convenience we alternated between solid and dotted lines.
The symbol $\Box$ denotes an upper limit. The times (JD$-$2442000) of the
spectra have been indicated. Negative spectral indices (e.g.\ day\,640,
644, 647) are indicative of optically thin synchrotron emission.
Rapid switches to positive spectral indices occur around day\,641, 645 and
648 and correspond to increases in the flux density at high
frequencies; following such spectral inversions the emission reverts
back towards an optically thin spectrum over a few days. These
spectral inversions are indicative of emission from new, initially
optically thick, components, which peak first at higher frequencies as
they expand (see text).}

\end{figure}

The light curves show that the decline is not a smooth power-law or
exponential decay.  It seems that there are various local maxima.
Especially the `newly' added measurement at 2380\,MHz near day\,640
obtained by Craft (1975) is far from that expected. It is substantially lower
than Owen et al.'s (1976) measurement near day\,641 at
2695\,MHz. This cannot be due to the slight difference in frequency.
So, a maximum was reached in the 2380/2695\,MHz band near Owen et
al.'s (1976) measurement (whether this is the main peak of the radio
outburst or an intermediate maximum we cannot say).  This is the
first time that it has been shown that there are multiple maxima in the
A0620$-$00 radio light curves. 

We also plotted the radio spectral evolution as a function of time
(Fig.~2). The spectra are drawn from data which were obtained within a
time span of maximum 0.36~days. At first, around day\,640 the data are
consistent with optically thin synchrotron emission, at least between
1.4--2.6\,GHz, with a spectral index 
($\alpha = \Delta \log S_{\nu} / \Delta \log \nu$) around $-$1. 
By the following day, i.e.\ day\,641, the radio
spectrum has inverted. Over the following three days, up until day\,644,
the spectrum slowly reverts towards an optically thin state. On day\,645
another inversion of the spectrum occurs. Possibly a third spectral
inversion also occurs on day\,648. We have indicated the times of the 
spectral inversions at the top of Fig.~1 as T2, T3, and T4?, respectively.
T1 corresponds to the start of the radio outburst which is not exactly 
known.

Such spectral changes, corresponding to local maxima in the light
curve which peak first at higher frequencies, are indicative of
repeated superposition of flares which are initially optically
thick. This effect is well observed in a sequence of five major radio
flares from Cyg\,X-3 in 1994 (Fender et al.\ 1997). Given that much
evidence points to the ejection of bright radio-emitting clumps from
Cyg\,X-3 which correspond to these flares (e.g.\ Geldzahler et al.\ 1983;
Mioduszewski et al.\ 1998), it seems natural to infer that the
secondary maxima and associated spectral changes that we see in the
radio light curve of A0620$-$00 correspond to multiple ejections of
synchrotron-emitting components from the source.

\subsection{Single baseline interferometry}

\begin{figure}
\centerline{\hbox{
\psfig{figure=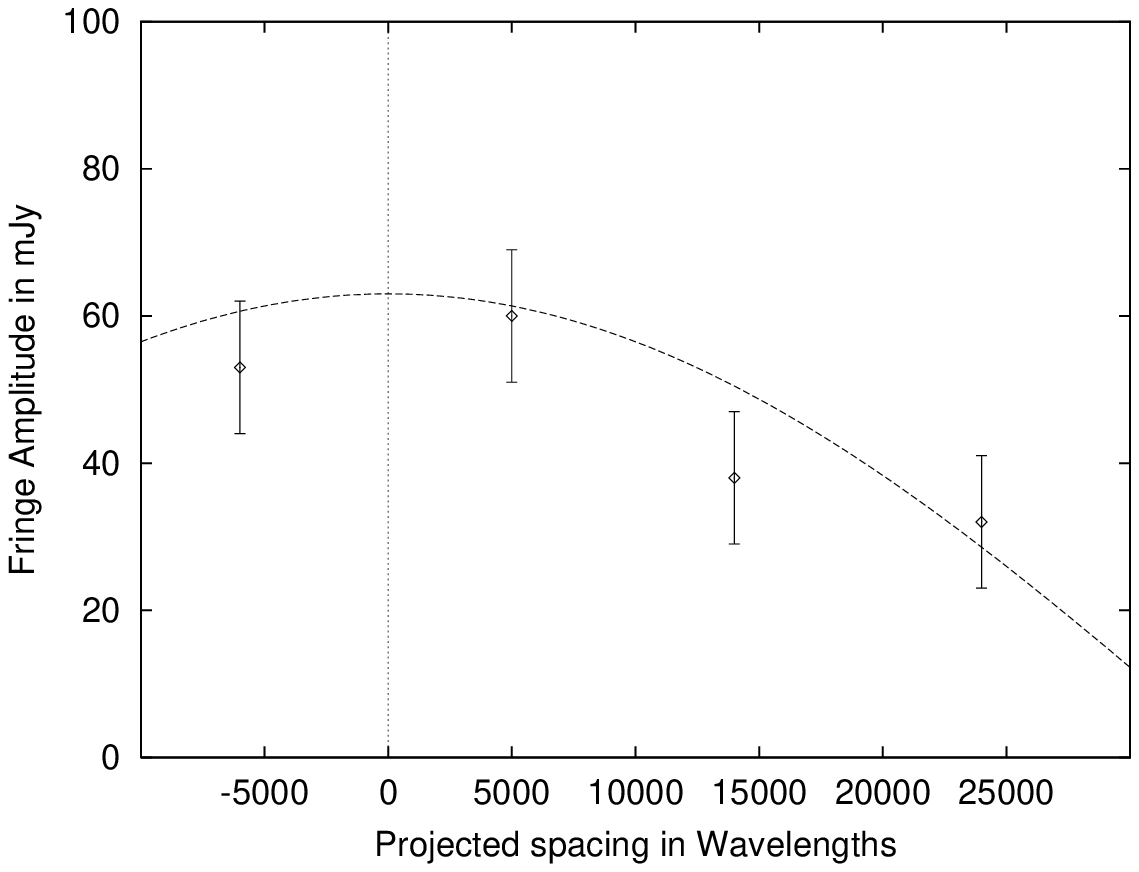,bbllx=50pt,bblly=30pt,bburx=410pt,bbury=292pt,clip=yes,width=9.0cm}}
}
\caption{Measured fringe amplitude on a single baseline versus (projected) 
baseline length from the 962\,MHz observations of A0620$-$00 on day\,648.
Fitting a model to the data enables structures to be found: a 3 arcsec 
double source fits reasonably well (solid line) 
and so does a gaussian of 4 arcsec fwhm.}
\end{figure}

Based on the suggestion that A0620$-$00 might exhibit jet ejections we
decided to re-examine the radio observations by Davis et al.\ (1975).
These observations were done with the MkII-MkIII interferometer 
(baseline 24\,km) at Jodrell Bank at 962\,MHz, where the resolution is 2.5
arcsec.  At that time no significant
variation with hour angle was detected during the observations.
The original raw data were recorded on paper
tape, now lost, but luckily various hand drawn plots were found. 
A reassesment of the errors by us show, however, that the measured fringe 
visibility varied systematically with
hour angle on 23rd August over 0800--1200 (UT), i.e.\
day\,647.8--648.0, in a way which suggested the presence of a slightly
extended source. The locus of interferometer baseline with hour angle forms
part of an ellipse in the resolution plane. The visibility curve 
(Fig.~3), though only sampled by taking 1 hour long integrations, was
similar to that of SS433, also measured by a single baseline
interferometer (Spencer 1979). 
The maximum amplitude occured at an hour angle which corresponds to a
position angle of 45$\pm$15 degrees of a slightly extended source. The
angular size of a source with simple structure can be found by projecting the
observed fringe amplitude onto a line through the origin of the resolution
plane at the position angle of the source. The figure shows the visibility
along this projected baseline which falls on both sides of the origin.  This
simple method of model fitting relies on the source being unresolved in a
direction perpendicular to the extension, which since the source is only
slightly resolved anyway, is a reasonable assumption.  We found that
A0620$-$00 was extended by
3--4~arcsec by fitting a double source or gaussian to the visibility
curve. We note that if the source was a point then there would not be
any significant variation of amplitude with baseline length.
The effective field of view of such an interferometer is only a few 
times the resolution and so the effects seen cannot be caused by 
confusion.
Further inspection of the NVSS does not reveal any obvious radio sources 
within 5 arcmin which may have been responsible for this apparent 
extension.
 
This observation was done on a date $\sim$20 days after the start of the 
X-ray outburst. If we assume that the extended source is a result from the
primary jet ejection and that this originated at the start of the 
X-ray outburst, the data imply an apparent expansion velocity of the jet 
of $\sim$0.9--1.2\,c, assuming a distance of 
1050\,pc (Shahbaz, Naylor \&\ Charles 1994). 
The apparent jet velocity, however, is $\sim$0.45--0.6\,c, if 
there was two-sided ejection in a pair of jets. Association of the
extended structure with later ejections we infer to have occurred
(Section 2.2) only increases this velocity.

\section{A comparison with similar X-ray transients}

\subsection{X-ray and radio light curves}

\begin{figure*}
\centerline{\hbox{
\psfig{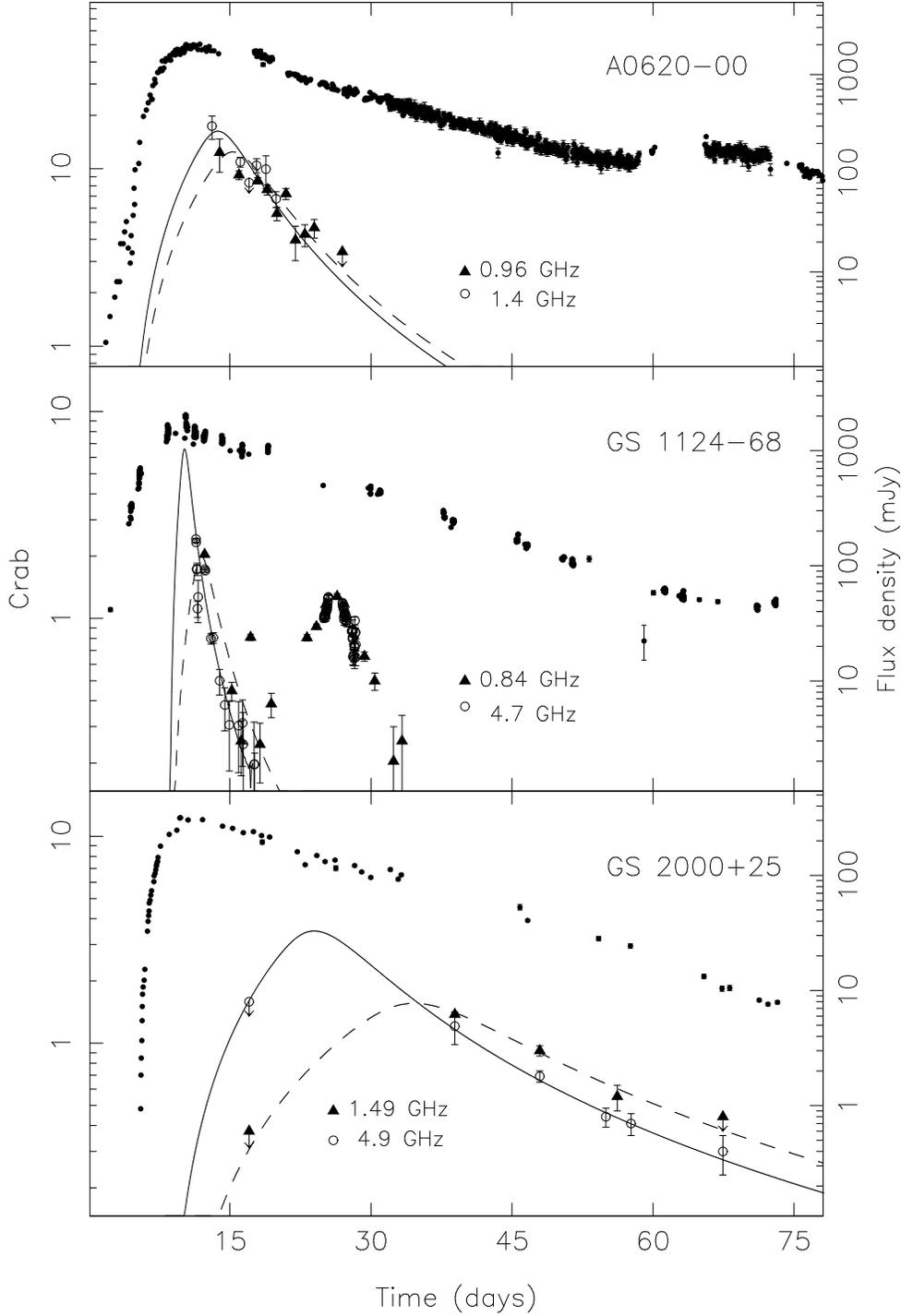}}
}
\caption{X-ray and radio light curves of A0620$-$00 (top), GS\,1124$-$68
(middle) and GS\,2000+25 (bottom). The X-ray measurements of A0620$-$00
are those obtained by the Ariel V SSE (2--18\,keV), Ariel V ASM (3--6\,keV) 
and \mbox{SAS-3} CSL-A (1.5--6\,keV), see Kuulkers (1998). The Ariel V SSE and ASM 
light curves have been scaled to match the \mbox{SAS-3} CSL-A light curve. The X-ray 
measurements of GS\,1124$-$68 are those obtained with the Ginga LAC
(1.2--37\,keV; Ebisawa et al.\ 1994) and the Ginga ASM (1--20\,keV; 
Kitamoto et al.\ 1992).
The rise of the X-ray outburst of GS\,2000+25 is taken from Tsunemi et al.\
(1989), while the rest of the light curve is from Kitamoto et al.\ (1992).
For the radio light curves we have plotted the measurements at two frequencies.
The radio measurements of A0620$-$00 are tabulated in Table 1, 
the GS\,1124$-$68 measurements are from Ball et al.\ (1995) and those of
GS\,2000+25 are obtained by Hjellming et al.\ (1988). Drawn with solid and
dashed lines are fits to the lowest and highest radio frequency
measurements, respectively, with single synchrotron bubble models as
presented by Hjellming et al.\ (1988) and Ball et al.\ (1995).}

\end{figure*}

\begin{figure*}
\centerline{\psfig{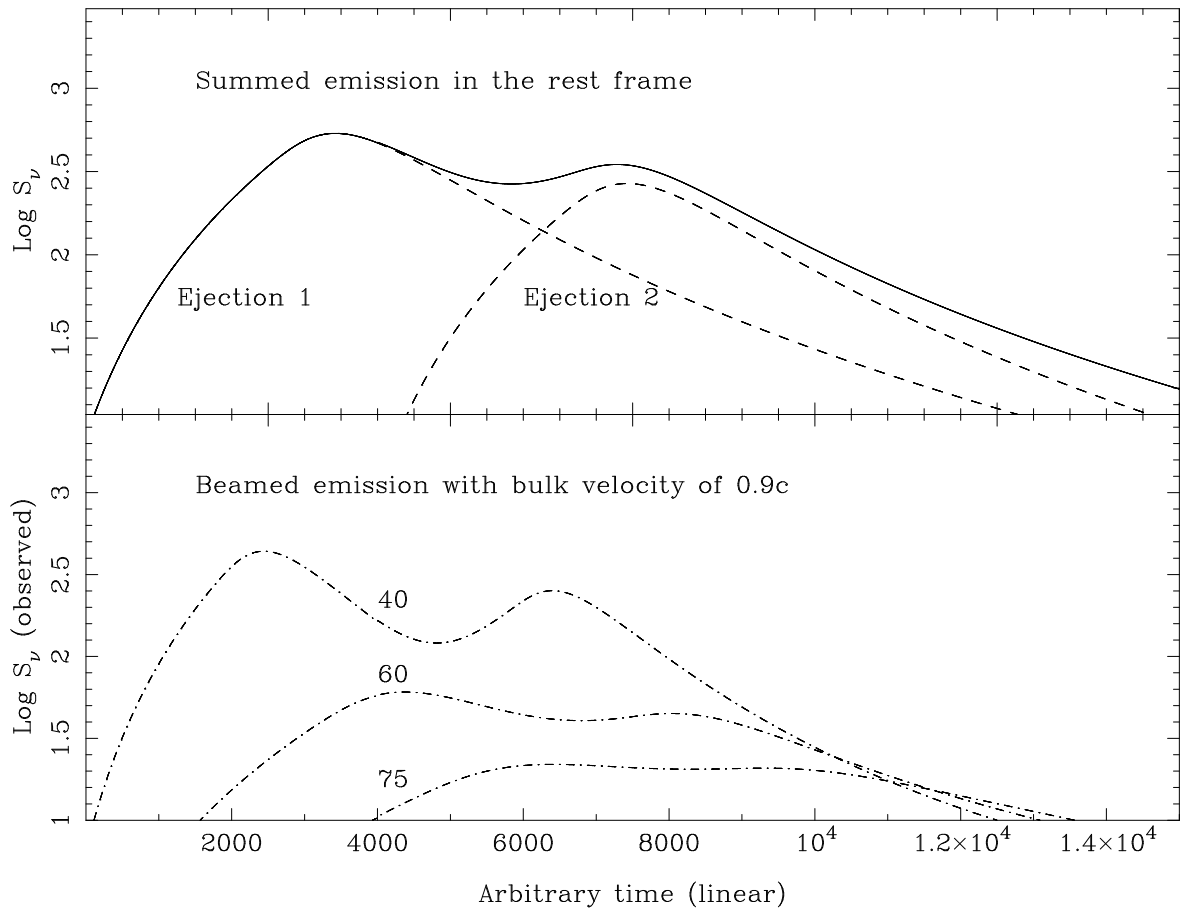}}
\caption{
Model radio light curves for observed emission from relativistic
outflows. The top panel shows the emission that would be observed in
the rest (comoving) frame of the ejecta; an ejection at $t=0$ is
followed by a second ejection, of half the amplitude, at
$t=4000$. The lower panel indicates the light curves that would be
observed if the ejecta are moving with a bulk velocity of $0.9c$ at
angles of 40, 60 and 75 degrees to the line of sight. The larger the
angle to the line of sight the weaker the radio emission and smoother
the light curve. In addition, for a fixed sensitivity limit emission
from outflows at large angles to the line of sight will appear to rise
later and last a little longer.
}
\end{figure*}

In Fig.~4 we show the X-ray light curves and radio light curves of
A0620$-$00 and two other black-hole X-ray transients with similar
X-ray light curves (see e.g.\ Chen et al.\ 1997), i.e.\
GS\,1124$-$68 and GS\,2000+25. 
For the radio we show the data at two frequencies in
order to get the longest and best coverage.  We have also indicated
the simple single synchrotron bubble model light curves as 
derived by Hjellming et al.\ (1988) and Ball et al.\ (1995).

As noted before (see Kuulkers 1998, and references therein), the X-ray light 
curve of A0620$-$00 shows an enhancement for a brief time just after the 
outburst peak. In the hard X-rays (6--15\,keV) this X-ray flare might be even
more pronounced (see Kuulkers 1998). 
Interestingly, the Ginga hard X-ray (9.3--37\,keV) light curve of 
GS\,1124$-$68 shows such a pronounced reflare just after the peak of the X-ray 
outburst (Ebisawa et al.\ 1994; see also Takizawa et al.\ 1997 and Fig.~6). 
A similar 
conclusion was drawn by Brandt et al.\ (1992) using Watch data. Note that
GRS\,1009$-$45 displays a similar hard X-ray feature (see Kuulkers 1998).
We suggest that this (hard) X-ray reflare is similar to the one seen in 
A0620$-$00. 

So, although the X-ray light curves are very similar in the three cases,
the radio light curves differ considerably. In particular, while we
have shown that there is evidence for multiple small ejections comprising
the A0620-00 light curve, it shows nothing like the major secondary
radio flare observed from GS 1124-68 (see also below).  The radio light
curve of GS\,2000+25 can not really be compared quantitatively with
those of A0620$-$00 and GS\,1124$-$68 since the parts of the X-ray
outbursts covered are different. However, it is interesting to note
that GS\,2000+25 could have experienced radio outbursts like
GS\,1124$-$68, since the covered parts of the X-ray outbursts
complement each other. Similarly, if GS\,1124$-$68 had been
covered longer it might have shown a similar decay as GS\,2000+25.

Fig.~4 shows that it is difficult to compare observed radio light
curves in order to infer similar characteristics, especially if the
coverage is different.  This applies even more when modeling such
light curves. Also, the start of the radio outbursts can not be determined,
since the very first rise has not been covered. The radio data are all
consistent with the start of the outburst being 
around the time of the start of the
X-ray outburst. Although modeling the radio light curves with a
synchrotron bubble model gives a start which lags the X-rays by about
ten days, it has been shown that such models do not describe the data
very well (see above).

The similar X-ray but dissimilar radio light curves for the three
X-ray transients are a strong indication that the radio
emission for all these sources arises in relativistic outflows. In
this case the X-ray emission is more or less isotropic and similar
behaviour is observed regardless of the inclination of the binary. On the
other hand, the radio emission, if it arises in relativistic outflows,
will be strongly beamed, with both brightness and morphology of light
curve affected by the angle to the line of sight. 
Our inferred jet velocities in the case of A0620$-$00 point to relativistic 
outflows.
%; we note that the report of $\sim$500\,keV line emission from 
%GS\,1124$-$68 (Goldwurm et al.\ 1992)
%also indicates the presence of energetic particles.

Fig.~5 shows simulated light curves for intrinsically identical radio
ejections viewed at differing angles to the line of sight. The model
is not meant to be a fit to the data but merely an indication of the
angle-dependence of the observed radio light curves. We assume a major
ejection at time $t=0$ followed by a secondary ejection of half the
initial amplitude, $\Delta t=4000$ later. The light curves are
superpositions of emission from approaching and receding components
from a symmetric ejection with a bulk velocity of 0.9\,c. Velocities
of this order have been inferred from observations of apparent
superluminal motions from GRS\,1915+105 (Mirabel \& Rodriguez 1994;
Fender et al.\ 1999) and GRO\,J1655$-$40 (Tingay et al.\ 1995; Hjellming \&
Rupen 1995). The individual ejections are modeled as simple
synchrotron bubbles using the formulation of Hjellming \& Johnston
(1988) in their rest frame.  We show the results for ejections at
angles of 40, 60 and 75 degrees to the line of sight (i.e.\ equally
distributed in $\cos i$). For an angle of 40 degrees to the line of
sight the emission is dominated by the emission from the two
approaching components and twin peaks are seen relatively close
together in time. As the angle increases the radio outburst is weaker,
broader and smoother. 

It is clear from these model light curves that if, as seems likely, the radio
emission from most, if not all, X-ray transients arises in
relativistic, collimated outflows, then we can expect a variety of
radio light curves depending on the angle the outflow makes to the
line of sight.  In general, broad, weak radio light curves
will indicate an outflow near to the plane of the sky
suffering Doppler de-boosting, as a result of the radiation being beamed 
away from the line of sight.  Sharper, brighter radio light curves
will indicate an outflow closer to the line of sight. In addition,
Fig.~5 shows that, for a given radio sensitivity limit we might expect
jets nearer to the plane of the sky to appear to rise later and last
(slightly) longer than those inclined nearer the line of sight.

Although our model light curves seem to match those observed at face value,
we note that the determined inclination of the three sources from 
optical/IR observations 
(e.g.\ Charles 1998) does not fully match from what we would infer
from the radio light curves. Ellipsoidal light curve modeling of 
the orbital light curves indicate that GS\,1124$-$68 has 
a {\em larger} inclination ($\sim$54 degrees) than A0620$-$00 
($\sim$37 degrees), whereas the radio 
light curve is more peaked in the case of GS\,1124$-$68 and thus one would 
infer a {\em smaller} inclination than compared to A0620$-$00.
Although we recognize this problem we note that other factors may  
complicate the light curves, such as the time between two ejections and 
their relative strength and their speed. At least we have shown that 
qualitatively radio light curves change as a function of inclination
when jets move near the speed of light, which may explain the observed radio 
behaviour.

\subsection{A closer look at GS\,1124$-$68}

\begin{figure}
\centerline{\hbox{
\psfig{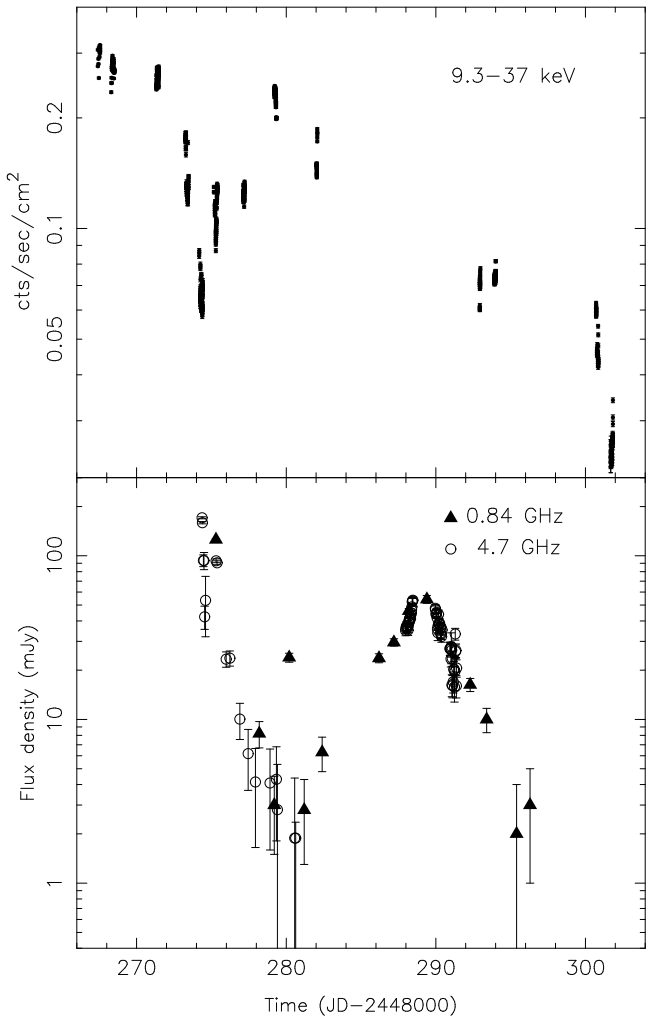}}
}
\caption{Ginga hard X-ray (9.3--37\,keV; Ebisawa et al.\ 1994) and radio
(843 and 4700\,MHz; Ball et al.\ 1995) light curves of GS\,1124$-$68 during
the beginning of the X-ray outburst.
}
\end{figure}

Brandt et al.\ (1992) concluded that the hard X-ray peaks 
in the outburst light curve of GS\,1124$-$68 were $\sim$13~days
apart, i.e.\ very similar to the time span between the first radio 
measurement and the peak of the radio reflare. However, according to 
Brandt et al.\ (1992) the radio was delayed by $\sim$7~days with respect to the
X-rays. In Fig.~5 we plot the Ginga hard X-ray (9.3--37\,keV) light curve 
and the radio light curve near the start of the outburst.
Indeed the correlation between the X-rays and radio is striking, but 
a detailed look reveals that a $\sim$7~day delay does not fit both
light curves. We conclude that the first rapid decline in the radio follows 
the rapid decline in the hard X-rays by about $\sim$3~days, whereas the peak 
of the second radio flare is $\sim$10~days after the second hard X-ray flare.
Such radio delays with respect to hard X-rays are not uncommon in 
X-ray transients; an example is GRO\,J1655$-$40 which has shown
similar hard X-ray and radio light curves near the beginning of its
outburst (e.g.\ Harmon et al.\ 1995).

\smallskip

\section{conclusions}

We have shown that the radio outburst of A0620$-$00 in 1975
is consistent with multiple ejection events, with the jets most probably
moving at near the speed of light, as has been inferred for
GRO\,J1655$-$40 and GRS\,1915+105. This strengthens the suggestion 
(e.g.\ Hjellming \&\ Rupen 1995) that most, maybe all, (soft) X-ray transients
undergo radio outbursts at the time of their X-ray and optical outbursts, and
that they generally consist of {\em multiple} ejections,
presumably of a significant fraction of the accretion disc.

We note that recently it has been reported that sources 
showing similar X-ray (spectral) behaviour as A0620$-$00 
(e.g.\ GS\,1124$-$68 and GS\,2000+25) do not contain 
a rapidly spinning black hole and it has been suggested that such 
systems can not form relativistic jets (Zhang, Cui \&\ Chen 1997; see also 
Cui, Zhang \&\ Chen 1998). 
Our single baseline interferometry suggests, however, relativistic
jet speeds in the case of A0620$-$00, 
which is in contradiction with their suggestion.
Determining whether black-holes spin or not from fairly simplistic X-ray
modelling may therefore be rather uncertain.

By comparing (soft) X-ray transients with similar X-ray behaviour
we find that the radio emission displays different light-curve shapes 
and strengths. This strongly supports isotropic X-ray emission, whereas
the radio emission is beamed (i.e.\ in the form of jets).
A first qualitative modeling of such a geometry seems to match the observed
variety of radio light curves, and seems to strengthen the hypothesis 
that the jets are moving at considerable speed. 

\section*{acknowledgments}

We gratefully acknowledge the use of the processed \mbox{SAS-3} data of A0620$-$00 from
Kenneth Plaks, Jonathan Woo and George Clark.
We thank Lewis Ball for providing the radio measurements of GS\,1124$-$68,
Wan Chen for the Ariel~V ASM data and part of the Ariel~V SSE data
of A0620$-$00, 
Ken Ebisawa for the Ginga LAC measurements of GS\,1124$-$68 and Shunji
Kitamoto for the Ginga ASM measurements of GS\,1124$-$68 and part of
GS\,2000+25.

\bsp % ``This paper has been produced using the ...''

\label{lastpage}


\begin{thebibliography}{}

\bibitem[]{}
Ball, L., 1994, ApJS, 90, 889
\bibitem[]{}
Ball, L., Kesteven, M.J., Campbell-Wilson, D., Turtle, A.J., Hjellming, R.M., 
1995, MNRAS, 273, 722
\bibitem[]{}
Bieging, J., Downes, D., 1975, Nat, 258, 307
\bibitem[]{}
Brandt, S., Castro-Tirado, A.J., Lund, N., Dremin, V., Lapshov, I., Sunyaev, R.,
1992, A\&A, 254, L39
\bibitem[]{}
Charles, P.A., 1998, in: Theory of Black Hole Accretion Disks, 
M.~Abramowicz, G.~Bjornsson \&\ J.~Pringle (eds.), CUP, p.~1
\bibitem[]{}
Chen, W., Shrader, C.R., Livio, M., 1997, ApJ, 491, 312
\bibitem[]{}
Condon, J.J., Cotton, W.D., Greisen, E.W., Yin, Q.F., Perley, R.A.,
Taylor, G.B., Broderick, J.J., 1998, AJ, 115, 1693
\bibitem[]{}
Craft, H.D., 1975, IAU Circ. 2822
\bibitem[]{}
Craft, H.D., Davis, M.M., 1998, private communication
\bibitem[]{}
Cui, W., Zhang, S.N., Chen, W., 1998, ApJ, 492, L53
\bibitem[]{}
Davis, R.J., Edwards, M.R., Morison, I., Spencer, R.E., 1975, Nat, 257, 659
\bibitem[]{}
Ebisawa, K., Ogawa, M., Aoki, T., Dotani, T., Takizawa, M., Tanaka, Y., Yoshida,
K., Miyamoto, S., Iga, S., Hayashida, K., Kitamoto, S., Terada, K., 1994, PASJ,
46, 375
\bibitem[]{}
Elvis, M., Page, C.G., Pounds, K.A., Ricketts, M.J., Turner, M.J.L., 1975, 
Nat, 257, 656
\bibitem[]{}
Feldman, P.A., 1975, IAU Circ. 2822
\bibitem[]{}
Fender, R.P., Bell Burnell, S.J., Waltman, E.B., Pooley, G.G., Ghigo, F.D.,
Foster, R.S., 1997, MNRAS, 288, 849
\bibitem[]{}
Fender, R.P., Garrington, S.T., McKay, D.J., Muxlow, T.W.B., Pooley,
G.G., Spencer, R.E., Stirling, A.M., Waltman, E.B., 1999, MNRAS, in press 
\bibitem[]{}
Geldzahler, B.J., Johnston, K.J., Spencer, J.H., Klepczynski, W.J., 
Josties, F.J., Angerhoffer, P.E., Florkowski, D.R., McCarthy, D.D.,
Matsakis, D.N., Hjellming, R.M., 1983, ApJ, 273, L65
%\bibitem[]{}
%Goldwurm, A., Ballet, J., Cordier, B., Paul, J., Bouchet, L., Roques, J.P.,
%Barret, D., Mandrou, P., Sunyaev, R., Chrazov, E., Gilfanov, M., Dyachkov, A.,
%Khavenson, N., Kovtunenko, V., Kremnev, R., Sukhanov, K., 1998, MNRAS, 389, L79
\bibitem[]{}
Hjellming, R.M., Han, X., 1995, in: Lewin, W.H.G., van Paradijs, J., 
van den Heuvel, E.P.J.\ (eds), X-ray Binaries, Cambridge, Cambridge 
University Press, p.~308
\bibitem[]{}
Hjellming, R.M., Johnston, K.J., 1988, ApJ, 328, 600
\bibitem[]{}
Hjellming, R.M., Rupen, M.P., 1995, Nat, 375, 464
\bibitem[]{}
Hjellming, R.M., Calovini, T.A., Han, X.H., C\'ordova, F.A., 1988, ApJ, 335, L75
\bibitem[]{}
Harmon, B.A., Wilson, C.A., Zhang, S.N., et al., 1995, Nat, 374, 703
\bibitem[]{}
Kitamoto, S., Tsunemi, H., Miyamoto, S., Hayashida, K., 1992, ApJ, 394, 609
\bibitem[]{}
Kuulkers, E., 1998, NewAR, 41, 1 (astro-ph/9805031)
\bibitem[]{}
Lequeux, J., 1975, IAU Circ. 2822
\bibitem[]{}
Little, A.G., Crawford, D.F., Murdoch, H.S., 1976, Nat, 261, 113
\bibitem[]{}
Mioduszewski, A.J., Hjellming, R.M., Rupen, , M.P., Waltman, E.B., 
Pooley, G.G., Ghigo, F.D., Fender, R.P., 1998, in: IAU~164 -- Radio 
emission from Galactic and Extragalactic compact sources, 
ASP Conf.\ Ser.\ 144, p.~351
\bibitem[]{}
Mirabel, I.F., Rodriguez, L.F., 1994, Nat 371, 46
\bibitem[]{}
Owen, F.N., Balonek, T.J., Dickey, J., Terzian, Y., Gottesman, S.T., 1976, ApJ,
203, L15
\bibitem[]{}
Scott, P.F., 1998, private communication
\bibitem[]{}
Scott, P.F., 1975, IAU Circ. 2823
\bibitem[]{}
Shahbaz, T., Naylor, T., Charles, P.A., 1994, MNRAS, 285, 607
\bibitem[]{}
Spencer, R.E., 1979, Nat 282, 483
\bibitem[]{}
Takizawa, M., Dotani, T., Mitsuda, K., et al., 1997, ApJ, 489, 272
\bibitem[]{}
Tingay, S.J., Jauncey, D.L., Preston, R.A., Reynolds, J.R., Meier, D.L.,
Murphy, D.W., Tzioumis, A.K., McKay, D.J., Kesteven, M.J., Lovell, J.E.J.,
Campbell-Wolson, D., Ellingsen, S.P., Gough, R., Hunstead, R.W., Jones, D.L.,
McCullough, P.M., Migenes, V., Quick, J., Sinclair, M.W., Smits, D.,
1995, Nat, 374, 141
\bibitem[]{}
Tsunemi, H., Kitamoto, S., Okamura, S., Roussel-Dupr\'e, 1989, ApJ, 337, L81
\bibitem[]{}
van der Laan, H., 1966, Nat, 211, 1131
\bibitem[]{}
Zhang, S.N., Cui, W., Chen, W., 1997, ApJ, 482, L155

\end{thebibliography}
\end{document}